# Evolution of magnetic anisotropy in cobalt film on nanopatterned silicon substrate studied in situ using MOKE

**Khushboo Bukharia[1], Prasanta Karmakar[2], Dileep Kumar[3], Ajay Gupta[1]***

[1]*Amity Centre for Spintronic Materials, Amity University, Noida 201313, India*
[2]*Variable Energy Cyclotron Centre, 1/AF, Bidhannagar, Kolkata 700064, India*
[3]*UGC-DAE CSR, University Campus, Khandwa Road, Indore 452001, India*

*\*Email: agupta2@amity.edu*

## Abstract

Evolution of magnetization behaviour of cobalt film on nano patterned silicon substrate, with film thickness, has been studied. *In situ* magneto-optical Kerr effect measurements during film deposition allowed us to study genuine thickness dependence of magnetization behaviour, all other parameters like surface topology, deposition conditions remaining invariant. The film exhibits uniaxial magnetic anisotropy, with its magnitude decreasing with increasing film thickness. Analysis shows that anisotropy has contributions from both, i) exchange energy which is volume dependent and, ii) stray dipolar fields at the surface/interface. This suggests that local magnetization follows only partially the topology of the rippled surface. As expected from energy considerations, for small film thickness, the local magnetization closely follows the surface contour of the ripples making the volume term as the dominant contribution. With increasing film thickness, the local magnetization gradually deviates from the local slope and approaches towards a uniform magnetization along the macroscopic film plane making the surface term as the dominant contribution. Significant deviation from the anisotropy energy expected on the basis of theoretical considerations can be attributed to several factors like, deviation of surface topology from an ideal sinusoidal wave, breaks of continuity along the ripple direction, defects like pattern dislocations, and possible decrease in surface modulation depth with increasing film thickness.

Anisotropic patterning of surfaces of magnetic films has been conventionally used to generate uniaxial magnetic anisotropy (UMA).[1,2] In a more controlled manner, uniaxial magnetic anisotropy can be generated by periodic nanopatterning of the surfaces of the substrate [3–6] or of the film itself.[7–9] Periodic nanopatterns can be produced in a variety of ways, for example, by i) ion erosion[3,10,11], ii) nano lithography[12,13] iii) appropriate annealing of sapphire substrate[14] etc. Ion beam erosion is a versatile technique by which nanoripples with controlled ripple wavelength and amplitude can be generated and this technique has been used in many studies to produce controlled UMA in thin films of a variety of materials like cobalt[3,4,15,16], Fe[17], permalloy[3,18] etc. Studies have been done on the variation of UMA with the wavelength and amplitude of the ripple[3,8] as well as with the thickness of the film.[16] Generally, the results are understood in terms of the model proposed by Schlomann[19] in which stray dipolar fields generated at the surface/interface are responsible for generation of uniaxial magnetic anisotropy. The anisotropy field is given by-

$$H_d = 4\pi M_s \frac{\pi \omega^2}{\lambda t}, \qquad \ldots(1)$$

Where $\omega$ is the *rms* roughness, $t$ is the thickness of the film, $\lambda$ is the ripple wavelength and $M_s$ is the bulk saturation magnetization. In case of film deposited on rippled substrate, both the interfaces of the film are rippled and contribute to the dipolar field/magnetic anisotropy and therefore, the total anisotropy field will be twice of that given by eq. 1.

Most of the studies in the literature have been done with the ripple amplitude in the range of a few nm where a good agreement with the Schlomann's formula[19] has been observed for sufficiently thick films.[8,9,16] From eq. 1 one can see that the anisotropy field should vary quadratically with the amplitude of ripples and thus, higher magnetic anisotropy can be obtained by increasing the ripple amplitude. Arranz *et.al.* addressed this aspect by studying the ripple amplitude dependence of anisotropy energy of Co film, the surface of which was patterned by ion erosion.[9] Only up to certain value of the parameter $\omega^2 / (\lambda t)$, the anisotropy energy follows Schlomann's formula, beyond which strong deviation from the theory are observed. The observed deviation has been attributed to possible discontinuity in the film as a result of ion erosion.

In order to understand various factors which may limit the maximum anisotropy induced by rippled surface topology, in the present work, we do a systematic study of the film thickness dependence of magnetic anisotropy of Co film on nanopatterned Si (100) substrate having ripple amplitude of 10 nm (peak to peak). Simultaneous measurement of the sheet resistance of the film provides some information about the growth behaviour of the film. *In-situ* measurements of both magneto-optical Kerr effect (MOKE) and sheet resistance inside the UHV chamber during film deposition ensure that all the parameters remain identical except the film thickness.

Nanorippled substrates were prepared by erosion of Si (100) substrate with $N_2^+$ ions of 5keV energy, incident at an angle of $60^0$ from the surface normal, using the ECR ion source at VECC, Kolkata.[20] The morphology of the irradiated substrate was investigated using Bruker Multimode Nanoscope V Atomic Forced Microscopy (AFM) apparatus. Deposition of Co film on nanopatterned substrate was done by electron beam evaporation in a UHV chamber which is equipped with facilities for doing *in-situ* resistance and MOKE measurements.[21] A quartz crystal thickness monitor was used to measure the film thickness. Resistance measurement was done using four probe technique in two directions – along the length of the ripples and normal to it. MOKE measurements were done in longitudinal geometry using a He-Ne laser. The base vacuum in the chamber was $1\times10^{-9}$ mbar. Resistance measurement was done continuously during film deposition. For MOKE measurement, deposition was stopped after depositing certain thickness and measurements were done along the length of the ripple and normal to it by rotating the sample.

Figure 1 gives the Atomic Force Microscopy (AFM) image of the nanopatterned Si substrate used for the deposition of the Co film. The average periodicity of the nanoripples was determined to be 60 nm and the average modulation depth of the ripple was 10.3 nm (peak to peak).

Figure 2 gives the sheet resistance of the Co film along and normal to the length of the ripples as a function of film thickness. One can see that around a thickness of 4 nm, the resistance drops down rapidly and beyond the thickness of about 4.5 nm, its variation become slower. Rapid decrease in resistance signals the coalescence of Co island to form a percolating cluster. From Fig. 2 one may also note that percolation along the length of the ripples takes place at a slightly smaller thickness relative to that along normal to the ripples. Thus, in conformity with earlier results, coalescence of Co island occurs preferentially along the length of the ripples.[4]

Figure 3(a) gives some representative *in-situ* MOKE hysteresis loops of Co films along and normal to the length of the ripples taken after deposition of different thicknesses. After deposition of 60 nm film, the sample was removed from the deposition chamber and detailed *ex situ* MOKE measurements were done as a function of azimuthal angle. Figure 3(b) shows the azimuthal angle dependence of remnant magnetization. It may be noted that the sample exhibits a uniaxial magnetic anisotropy having its easy axis along the length of the ripples.

The magnitude of magnetic anisotropy energy was calculated by taking the slope of hysteresis loop along the hard direction in the region of small fields.[15,22] It may be noted that the remanence does not become zero even along the hard axis, suggesting that there is some distribution in the direction of hard axis.[23,24] However, the azimuthal angle dependence of remanence could be very well fitted with Stoner-Wohlfarth model (results not shown) giving the value of the orientation ratio as 1.5.[24] The saturation magnetization was calculated by taking the known value of the magnetic moment of the Co as $1.72\mu_B$ per atom. In general, the total anisotropy energy per unit volume will have contributions from both volume as well as surface anisotropies and can be written as:

$$K_u^t = k_u^v + \frac{k_u^s}{t}, \qquad \ldots (2)$$

where $K_u^t$ is the total uniaxial magnetic anisotropy per unit volume, $k_u^v$ is the volume contribution, $k_u^s$ is the surface contribution and *t* is the thickness of the film. Therefore a plot of $K_u^t \times t$ *vs t* would give a straight line with its slope being the volume anisotropy $k_u^v$ and the intercept being the surface term $k_u^s$. Figure 4(a) gives the plot of $K_u^t \times t$ *vs t* for the present data. One may note that the curve is not a straight line rather, with increasing film thickness, the slope and hence the volume contribution exhibits a decrease. The t dependence of $K_u^t \times t$ was fitted with a polynomial, differential of which gives the local slope. The data could be fitted with a polynomial of 2$^{nd}$ order:

$$K_u^t \times t = a + bt + ct^2, \qquad \ldots (3)$$

*with, $a = 0.030 \pm 0.008$, $b = 0.0158 \pm 0.0006$, $c = (-1.15 \pm 0.09) \times 10^{-4}$.* Therefore, the volume contribution to the anisotropy, which is proportional to the local slope, would decrease linearly with increasing film thickness.

Liedke *et.al.* studied variation of magnetic anisotropies of ferromagnetic metals on nanoripples Si (100) substrate with film thickness.[16] They could identify two regimes of film thickness: i) For smaller thicknesses, the local spin follows the contour of the rippled surface and thus, there was no surface contribution to anisotropy due to stray dipolar fields and the anisotropy originated mainly due to exchange interaction in the bulk of the film. ii) For higher thicknesses, they observed sudden transition to a region in which the magnetization attains a uniform state and the anisotropy is purely of dipolar origin and could be explained in terms of the formula derived by Schlomann.[19] Chen *et.al.* studied ultra-thin films of cobalt in thickness range of a few monolayers on nanorippled MgO (001) substrate.[15] They could explain the thickness dependent magnetic anisotropy in terms of a volume contribution alone, though they also observed a negative surface contribution of unknown origin. A good quantitative agreement was obtained for the volume term with the theoretical expression for exchange energy derived assuming a sinusoidal variation of surface topology. Thus, one can see that in the extreme case of large ripple amplitude and small film thickness, the magnetization perfectly follows the surface contour and magnetic anisotropy has only volume contribution, with no contribution from possible stray dipolar fields.[15] In the other extreme case of small ripple amplitude and large film thickness, all the spin gets align parallel to each other and the anisotropy has its origin mainly in the stray dipolar field generated at the surface/ interface. One expects that in the intermediate case of large ripple amplitude as well as large film thickness, one should have an intermediate situation where the spin follows only partially the surface contour as depicted schematically in Fig. 5. In this case, both the stray dipolar field (surface contribution) and the exchange term (volume contribution) should contribute to the total anisotropy. This happens to be the situation in the present case. In addition, in case some grain texture develops during film growth that would also contribute to the volume anisotropy. However, in an earlier study it was shown that Co deposited on a nanorippled Si substrate develops only a very weak grain texture and the associated magnetic anisotropy is negligibly small as compared to the total morphology-induced magnetic anisotropy.[4] Following Chen *et.al*[15], one can express surface morphology as,

$$h_s(x) = h \sin\left(2\pi \frac{x}{\lambda}\right), \qquad \ldots (4)$$

where *h* and *λ* are the amplitude and wavelength of the rippled structure. In case when the local magnetization follows only partially the surface contour, one can write the *x* dependence of the local magnetization direction as,

$$h_m(x) = h' \sin\left(2\pi \frac{x}{\lambda}\right), \qquad \ldots (5)$$

Where *h'< h*. *h'=h* would mean that the local magnetization perfectly follows the surface contour/corrugation while, *h'=0* corresponds to a situation where magnetization is parallel to the macroscopic film plane.

In this case, following Chen *et.al.*[15], the exchange energy equation can be written as,

$$<f_{ex}> = \frac{A}{2}\left(\frac{2\pi}{\lambda}\right)^2 \left(\frac{2\pi h'}{\lambda}\right)^2 \left[\frac{1}{\sqrt{1+\left(\frac{2\pi h'}{\lambda}\right)^2}}\right], \qquad \ldots (6)$$

where *A* is the exchange stiffness constant.

as one can see from Fig. 4(a), the volume contribution to the total anisotropy decreases monotonically with increasing film thickness which would imply that parameter *h'* decreases with film thickness. Figure 4(b) gives the variation of *h'* with film thickness as calculated using eq. 6. One can see that for the smallest film thickness of 5 nm, *h'* is 4.5 nm which is close to the ripple amplitude of 5 nm. As expected with increasing film thickness, the amplitude of the variation of local magnetization direction keeps on decreasing and for the highest thickness of 60 nm, *h'* is only 1.6 nm.

For calculating the surface contribution, one may note that the dipolar charges generated at the surface will be proportional to the angle between the spin direction and the local surface slope and thus proportional to *h- h'*. Therefore surface contribution to the anisotropy energy can be obtained using Schlomann's formula[19] for demagnetizing field by replacing *h* by *h-h'*:

$$H_d = 4\pi M_s \frac{\pi(h-h')^2}{\lambda t}, \qquad \ldots (7)$$

Figure 4(c) gives the experimental surface contribution as obtained by subtracting the volume term from the total anisotropy energy and the corresponding theoretical value of surface energy as obtained using eq. 7. The theoretical surface energy was obtained by converting demagnetizing field calculated using Schlomann's formula. One can see that the experimental surface contribution is significantly lower than that calculated using above model. This large discrepancy can be understood in terms of several factors: i) with increasing film thickness the surface topography of the film is expected to get smeared out resulting in decrease in the modulation depth h, ii) the actual surface/interface contour of the film deviates from an ideal sinusoidal shape in terms of various defects like height corrugation and breaks of continuity along the ripple direction and overlapping ripples and defects like pattern dislocation.[8]

In conclusion, evolution of magnetic anisotropy of cobalt film deposited at normal incident on nanopatterned silicon substrate has been studied with increasing film thickness. *In-situ* MOKE measurements during film deposition allowed us to follow the magnetization behaviour with film thickness, keeping all the other parameters like surface topology and deposition conditions identical. Film exhibits well defined uniaxial magnetic anisotropy with easy axis along the length of the ripples. Thickness dependence of anisotropy energy, as obtained from hysteresis curve, has been analysed in terms of volume and surface contributions. The observed results can be understood in terms of a model in which the local magnetization only partially follows the surface topology, thus, resulting in both volume and surface contributions to the magnetic anisotropy. It is found that while for the smaller film thickness, the local magnetization almost completely follows the surface contour of the ripples; with increasing film thickness, it gradually deviates from the local slope, moving towards a uniform magnetization along the macroscopic film plane. It is found that the surface term is significantly smaller than that expected on the basis of Schlomann's formula. This deviation can be attributed to the deviations of surface topology from an ideal sinusoidal shape in terms of various defects like height corrugation and breaks of continuity along the ripple direction and defects like pattern dislocation.

The present results have important implication on the possible control of magnetic anisotropy by varying the ripple amplitude and wavelength. One expects that with increasing ripple amplitude, the anisotropy energy would not increase as $h^2$, since with increasing $h$, the local magnetization will gradually start deviating from the surface contour, thus resulting in a deviation from the value as expected based on Schlomann's model. Further, deviations from the ideal sinusoidal variation of the surface contour would also result in decrease in the anisotropy energy, as also observed in some earlier cases.[8]

Authors are thankful to Mr. Lavanya Kumar and Mr. Dipak Bhowmik of VECC, Kolkata for their kind support during ion irradiation and AFM measurements, respectively. Ms. Khushboo Bukharia is Junior Research Fellow supported by UGC-DAE CSR, Indore. Financial support from BRNS through project no. 37(3)/14/17/2015/BRNS is thankfully acknowledged.

**Figure Caption-**

Figure 1: (a) AFM image of nanorippled Si substrate used for deposition of Co film. The inset shows the fourier transformation of rippled pattern (b) line profile showing wavelength and modulation depth of the ripple.

Figure 2: Total sheet resistance R, measured as a function of film thickness, i) along the length of the ripples (continuous line), and ii) normal to it (dotted line).

Figure 3: (a) Some representative magnetic hysteresis loops of Co film along the easy axis (continuous line) and ii) hard axis direction (dotted line), (b) Remnant magnetization as a function of azimuthal angle for Co film of thickness 60nm.

Figure 4: (a) Thickness dependence of magnetic anisotropy energy of Co film, (b) Plot of h' as a function of film thickness, (c) Experimental surface contribution and calculated surface contribution based on Schlomann's formalism, as a function of film thickness.

Figure 5: Schematic diagram showing variation of local magnetization direction marked by arrows, relative to the surface/interface contour of the film.

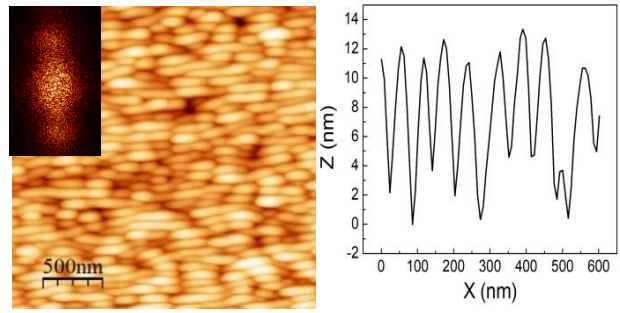

(a)   (b)

Fig 1

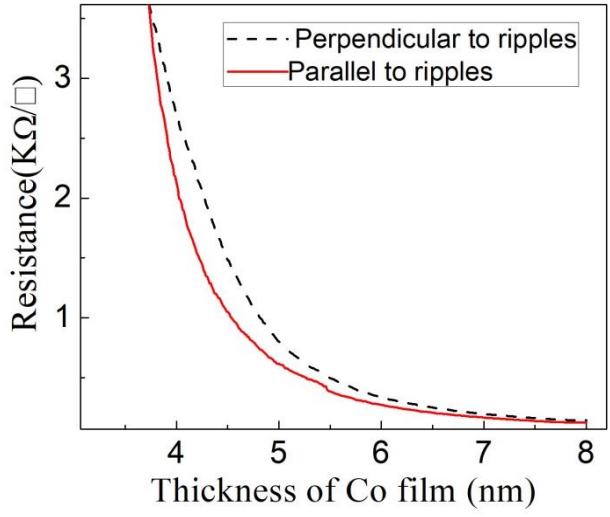

Fig 2

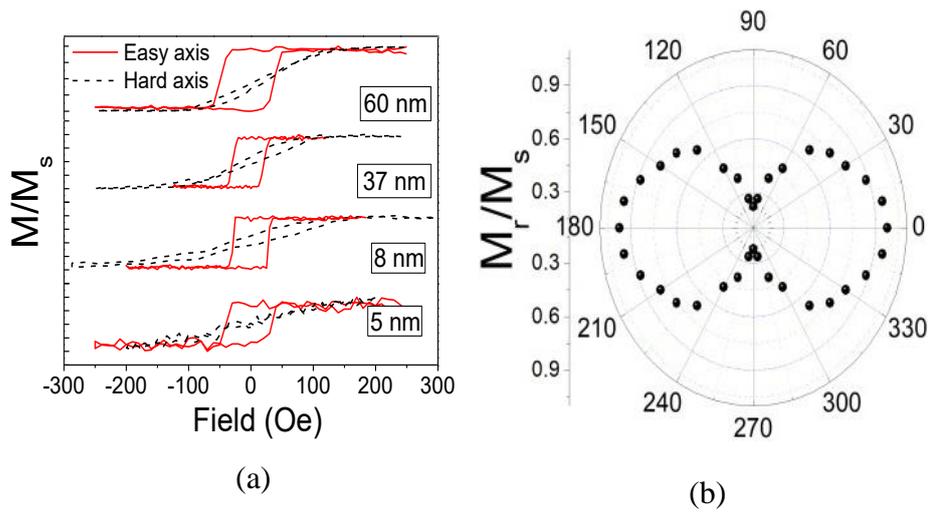

Fig 3

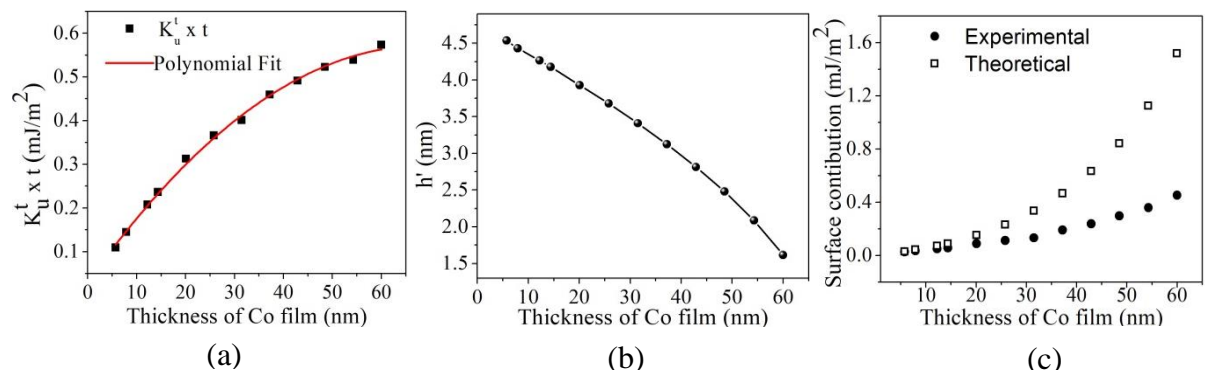

Fig 4

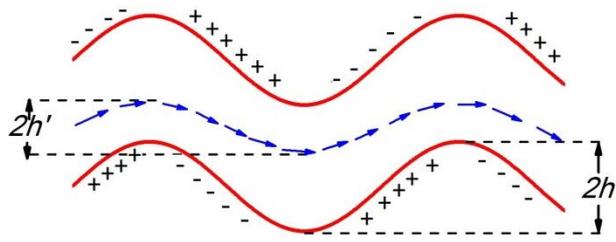

Fig. 5